\documentstyle[12pt,a4,epsfig]{article}
\begin{document}

\begin{flushright}
FIAN/TD-98/24
\end{flushright}

\medskip

\begin{center}
{\bf MINIJET TRANSVERSE ENERGY PRODUCTION IN THE NEXT-TO-LEADING
 ORDER IN HADRON AND NUCLEAR COLLISIONS}
\end{center}
\medskip
\begin{center}
{\bf Andrei Leonidov and Dmitry Ostrovsky}
\end{center}
\medskip
\begin{center}
{\it P.N.~Lebedev Physics Institute, 117924 Leninsky pr. 53, Moscow, Russia}
\end{center}
\bigskip
\begin{center}
{\bf Abstract}
\end{center}
\medskip
The transverse energy flow generated by minijets in hadron and nuclear
collisions into a given rapidity window in the central region is calculated in
the next-to-leading (NLO) order in QCD at RHIC and LHC energies. The NLO
transverse energy production in pp collisions
cross sections are larger than the LO ones by the factors of $K_{RHIC} \sim 1.9$
and $K_{LHC} \sim 2.1$ at RHIC and LHC energies correspondingly. These results
were then used to calculate transverse energy spectrum in nuclear collisions in
a Glauber geometrical model. We show that accounting for NLO corrections in the
elementary pp collisions leads to a substantial broadening of the $E_{\perp}$
distribution for the nuclear ones, while its form remains practically unchanged.

\newpage

\section{Introduction}
Minijet physics is one of the most promising applications of perturbative QCD to
the analysis of processes with multiparticle production. The minijet approach
is based on the fact that some portion of transverse energy is produced in
the semihard form, i.e. is perturbatively calculable because of the relatively
large transverse momenta involved in the scattering but is not observed
in the form of
customary hard jets well separated from the soft background.
A notable
feature of this approach is a predicted rapid growth of
 the integrated perturbative
cross section of
parton-parton scattering, responsible for perturbative transverse energy
production, with energy, and at RHIC
(200 GeV/A in CMS) and especially LHC (5500 GeV/A) energies
the perturbative cross section becomes quite large and in fact  even exceeds
the inelastic and total cross sections for large enough rapidity intervals.
The crossover from the regime described
by conventional leading twist QCD and the one where multiple hard interactions
are important is one of the most important problems of the minijet approach
\cite{JL}. The field is actively developing,
 recent reviews on the subject
containing a large number of references are  e.g. \cite{XNW} and \cite{KE1}.

    A special importance of minijet physics for ultrarelativistic heavy
ion collisions is due to the fact that minijets with large enough
transverse momenta are produced at a very early stage of the collision
thus forming an initial parton system that can further evolve kinetically
or even hydrodynamically, so that the minijet physics describes the initial
conditions for subsequent collective evolution of parton matter
\cite{BM}, \cite{KLL}, \cite{EKL}, \cite{EKR1},
see also a recent review \cite{GG}.

    Among the recent developments let us mention a new approach to minijet
production based on the quasiclassical treatment of nuclear gluon
distributions, \cite{KMW}, \cite{KR}, \cite{MMR}, \cite{GM}, \cite{KV} and
a description based on the parton cascade approach \cite{M}.

 The perspective of having a
perturbatively controllable description of a substantial part of the inelastic
cross section is certainly very exciting \footnote{The mechanism responsible for
the growth of the inelastic cross section as such can be soft, see \cite{JL}
and the recent analysis \cite{MW}.}
However, to determine the accuracy
of the predictions given by minijet physics, the existing calculations have
to be expanded in several directions.

   In this paper we discuss a conceptually simplest extension of the
leading order (LO) calculation of the transverse energy spectrum produced
 in heavy ion collisions presented in  \cite{EKL}, \cite{EKR} and
 \cite{EK} by including the  next-to-leading order (NLO) contributions
 to this cross section.
The NLO corrections to conventional high $p_{\perp}$ jet production
cross section  were
computed in \cite{EKS}, \cite{KS} and \cite{AGCG}
for Tevatron energies. Later the
code of \cite{KS} was used for calculating this cross section for RHIC and
LHC energies in \cite{EW}.
 The necessity of doing this computation in the minijet region was, of course,
 clearly understood and emphasized in the literature on minijet physics
\cite{KE1}, \cite{KE2}.
 A clear goal here is to establish an accuracy of the
LO prediction by explicitly computing the NLO corrections to it.

The outline of the paper is as follows.

 In the second section we describe a
calculation of the next-to-leading order (NLO) cross section of transverse
energy  production in proton-proton collisions
using the Monte Carlo code developed by Z.~Kunzst and
D.~Soper \cite{KS} and study a deviation from the LO result.

 In the third section the computed NLO cross section is used in the calculation
of transverse energy production in heavy ion collisions, where the nuclear
collision is described as a superposition of the nucleon-nucleon ones in a
Glauber geometrical approach of \cite{EKL}. We show that NLO corrections lead
to a substantial broadening of the transverse energy spectrum.

In the last section  we discuss the  results and formulate the conclusions.

\section{NLO minijet transverse energy production in hadron collisions}

The basic difference of  minijet physics from that of the usual high-$p_t$
jets is that the minijets can not be observed as jets as such. Experimentally
they
manifest themselves in more inclusive quantities such as the energy flow into a
given rapidity window. The NLO calculation of a jet cross section contains a
so-called jet defining algorithm specifying the resolution for the jet to be
observed, e.g. the angular size of the jet-defining  cone, see e.g. \cite{S}.
 As a  minijet can not be observed as
an energy flow into a cone separated from the rest of the particles produced in
the collision, some of the restrictions employed in the standard definition
of a jet should be relaxed. A natural idea is to define a
minijet produced "jet" as a transverse energy
deposited in some (central) rapidity window and a full azimuth.
This corresponds  to a
standard detector setup for studying central rapidity region in
nuclear collisions at RHIC and LHC.

The inclusive cross section is obtained by integrating the differential one over
the phase space on the surface corresponding to the variable in question.
Schematically the NLO distribution of the transverse energy produced into a
given rapidity interval $y_a < y < y_b$ is equal to
\begin{eqnarray}
\frac{d\sigma}{dE_\perp}=\int D^2PS  \frac{d\sigma}{d^4p_1d^4p_2}
\delta(E_\perp-\sum\limits_{i=1}^2 |p_{\perp
i}|\theta(y_{min}<y_i<y_{max}))\nonumber\\
+\int D^3PS  \frac{d\sigma}{d^4p_1d^4p_2d^4p_3}
\delta(E_\perp-\sum\limits_{i=1}^3 |p_{\perp i}|\theta(y_{min}<y_i<y_{max}))
\end{eqnarray}
 where the first contribution corresponds to the two-particle
final state and the second contribution  to the three-particle one. The
two-particle contribution has to be computed with one-loop corrections taken
into account.

  As in all NLO calculations the separation between the real emission and
virtual exchange requires regularization, in addition to the usual ultraviolet
renormalization, of the infrared and collinear divergences.
Physically this means that a distinction between a two-particle
and three-particle final state becomes purely conventional.

 In perturbative QCD one can meaningfully compute only infrared stable
quantities \cite{SW}, in which the divergences originating from real and virtual
gluon emission cancel each other, so that adding very soft gluons does not
change the answer.
  It is easy to convince oneself, that the
transverse energy distribution into a given rapidity interval Eq.~(1)
 satisfies the above requirement.

In more physical terms the main difference between the LO and NLO calculations
is that unlike the LO case the number of produced (mini)jets is no longer an
infrared-stable quantity in the NLO computation,
 i.e. it can not be predicted. For example, we can no
longer calculate a probability of hitting the acceptance window by a
given number of minijets, which is one of the important issues in an
event-by-event analysis of the initially produced gluon system (for details see
\cite{LO}). In view of the applications of the NLO results for nucleon-nucleon
collisions to the nuclear ones this means in turn that the elementary block in
the geometric approach is no longer describing the  production of a fixed number
of jets (2 jets in the LO case), but the production of transverse energy into a
kinematical domain specified by the jet defining algorithm.

\begin{figure}
 \begin{center}
 \epsfig{file=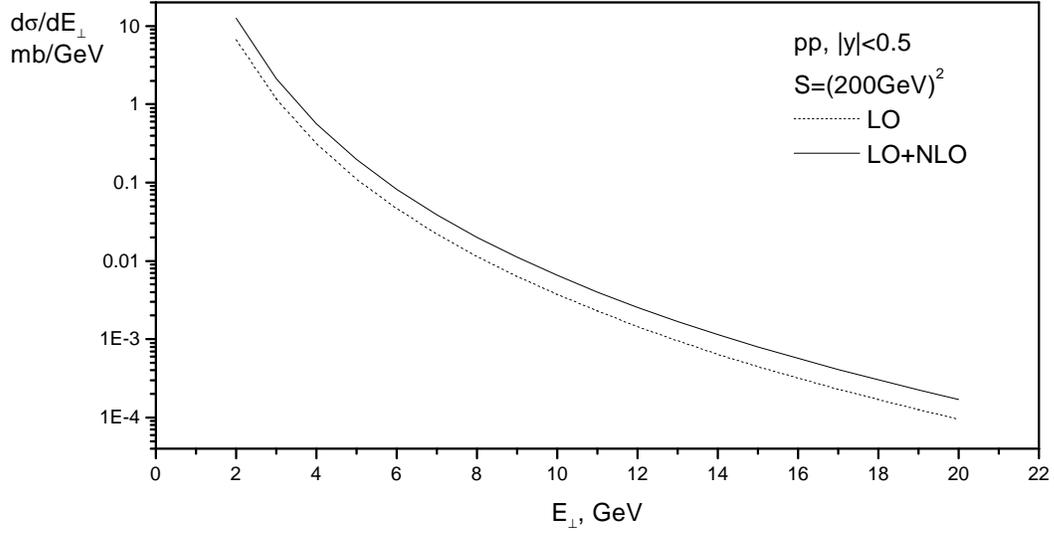,width=14cm}
 \end{center}
 \caption{NLO
 (solid line) and LO (dashed line) transverse energy spectrum in a unit central
 rapidity window for pp collisions at RHIC energy $\sqrt{s}=200$ GeV}
 \label{pp200}
\end{figure}

\begin{figure}
 \begin{center}
 \epsfig{file=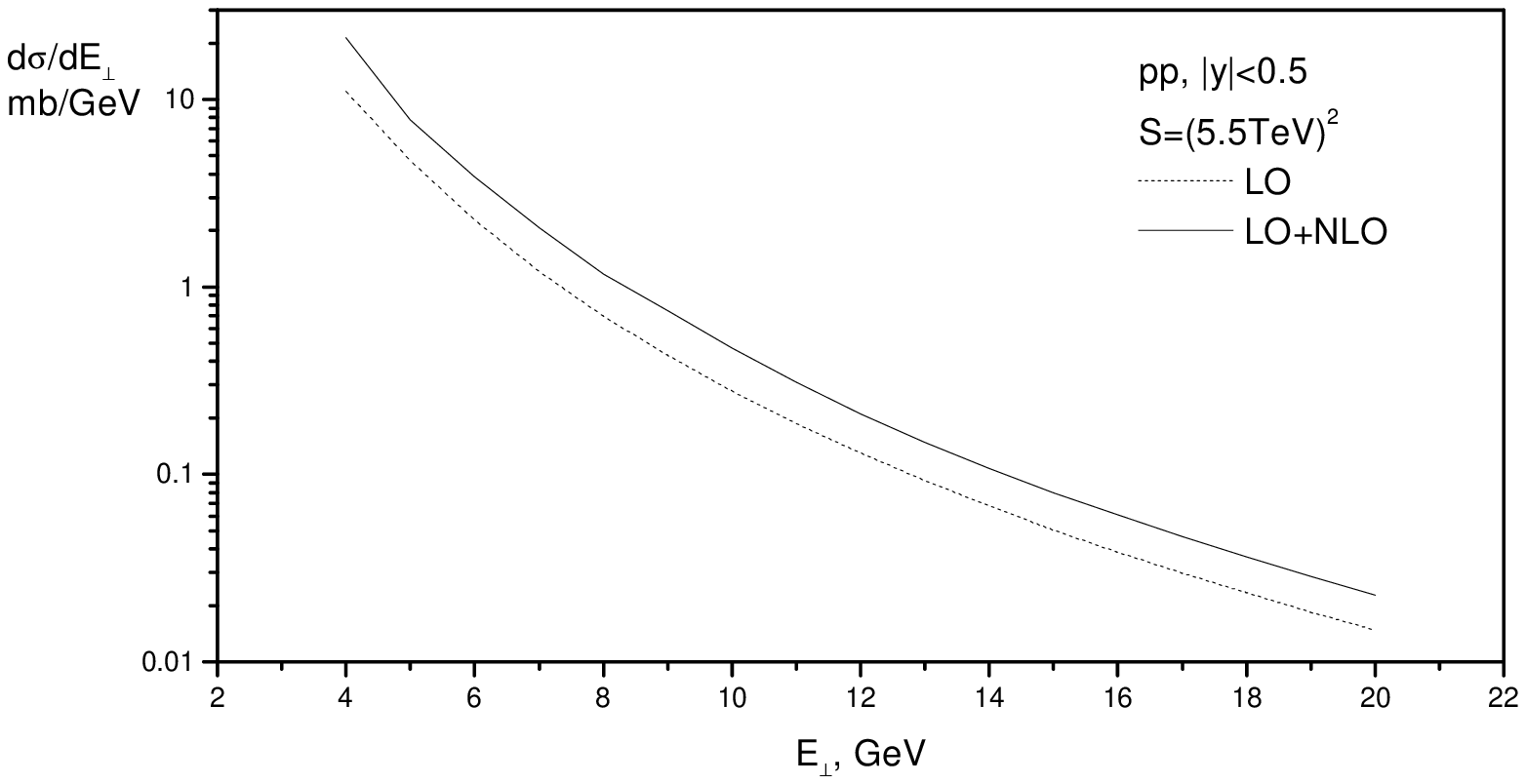,width=14cm}
 \end{center}
 \caption{NLO (solid line) and LO (dashed line) transverse energy spectrum in a
 unit central rapidity window for pp collisions at LHC energy $\sqrt{s}=5500$
 GeV}
 \label{pp5500}
\end{figure}

The calculation  of  transverse energy spectrum was performed using the
Monte-Carlo code developed by Kunzst and Soper \cite{KS}, with  a  jet
definition appropriate for transverse energy production specified in Eq.~(1).
The results for the cross section of transverse energy production into a central
rapidity window $ -0.5 \le y \le 0.5$ are presented in Fig.~\ref{pp200}
for RHIC energy  ($\sqrt{s}=200$ GeV) and Fig.~\ref{pp5500} for
the LHC one ($\sqrt{s} = 5500$ GeV), where for LHC we
have chosen the energy of proton - proton collisions available for protons in
the lead nuclei in PbPb collisions.
 The calculaions were performed using the MRSG(-)
structure functions \cite{MRSG}. The parameters for the fits for the
spectra having the functional form $a*E_{\perp}^{-\alpha}$ are given in Table 1.

\bigskip

\noindent
\begin{tabular}{|r|l|c|c|r@{.}l|c|c|}
\hline
\rule{0pt}{12pt}$\sqrt{S}$, GeV & LO/NLO & $\alpha$ & $a$, mb/GeV    &%
\multicolumn{2}{|c|}{$\sigma(E_0)$, mb} &%
$\sigma_{in}$, mb & $E^*_0$, GeV\\
\hline
   5500  & LO        & 4.14     & $3.7\cdot10^3$ &\quad 15&3  & 66.3 &     2.5 \\
\hline
   5500  & LO+NLO    & 4.24     & $7.8\cdot10^3$ &\quad 26&9  & 66.3 &     3.0 \\
\hline
    200  & LO        & 4.91     & $2.7\cdot10^2$ &\quad  0&31 & 41.8 &     1.1 \\
\hline
    200  & LO+NLO    & 4.92     & $5.0\cdot10^2$ &\quad  0&55 & 41.8 &     1.3 \\
\hline
\end{tabular}

\begin{center}
\bf Table 1
\end{center}

We see that taking into account the NLO corrections to transverse energy
production can roughly be described by introducing effective $K$-factors
$K_{RHIC} \sim 1.9$ and $K_{LHC} \sim 2.1$.
This agrees well with the "expected" $K$-factor used in earlier publications
\cite{XNW}, \cite{KE1}.
Let us note, that while at RHIC
energies the slopes of the LO and NLO curves are practically the same, at LHC
energies the NLO slope is about 2 percent larger than the LO one.

  In the third column we give the values of the integrated minijet production
cross section
\begin{equation}\label{sE0}
\sigma(E_0) = \int_{E_0}^{\infty} d E_{\perp}\ {d \sigma \over {d E_{\perp}}}
 \end{equation}
for the cutoff value $E_0 = 4$ GeV. The range of validity of a leading twist
approximation for transverse energy production in any given rapidity window is
determined by the inequality
\begin{equation}
 \sigma(E_0) = \int_{E_0}^{\infty} d E_{\perp}\ {d \sigma \over
{d E_{\perp}}} \leq \sigma_{inel}
\end{equation}
The equality in the above formula fixes the
limiting value of the cutoff $E_0^*$. The values of the inelastic cross section
are shown in the fourth column of Table 1%
\footnote{The inelastic cross section is computed using the parameterization
$\sigma_{inel}(s)=\sigma_0*(s/s_0)^{0.0845}*(0.96-0.03*\log (s/s_0))$, where
$s_0=1$ GeV, $\sigma_0=21.4$ mb, which gives a good description of the existing
experimental data \cite{exp}}
 and the limiting cutoff values $E_0^*$ are
shown in the fifth one. We see that the limiting values $E^*_0$ are quite small.
Let us stress hat the values of $E_0^*$
 depend rather strongly  on the rapidity window under
consideration. The limiting cutoff values shown in Table 1 refer to the central
unit rapidity interval and thus present a lower bound with respect to those
corresponding to larger rapidity intervals.

  Let us also note that, as mentioned before, we had to fix a scale
for the NLO logarithmic corrections, which for the above calculations was chosen
to be $\mu = E_{\perp}$.
 We have checked that variations of this scale in the range
$0.5\ E_{\perp} \le \mu \le 1.5\ E_{\perp}$ does not produce significant
variations of the result, so that the NLO calculation is stable and thus
produces a reliable prediction for the difference between the LO and NLO cases.

\section{NLO transverse energy production in nuclear collisions}

\begin{figure}[h]
 \begin{center}
 \epsfig{file=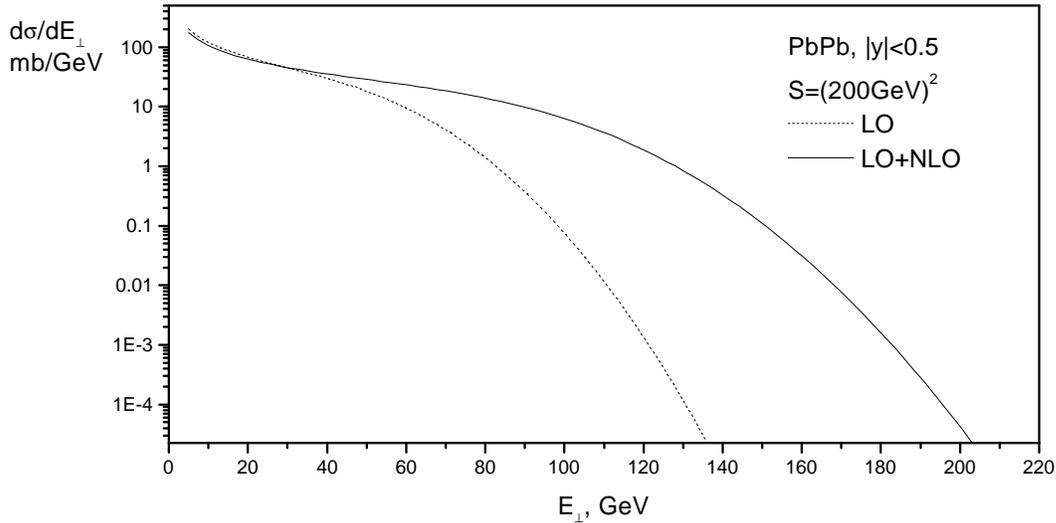,width=14cm}
 \end{center}
 \caption{NLO (solid line) and LO (dashed line) transverse energy spectrum in a
unit central rapidity window for PbPb collisions at RHIC energy $\sqrt{s}=200$
GeV}
 \label{PbPb200}
\end{figure}

\begin{figure}
\begin{center}
\epsfig{file=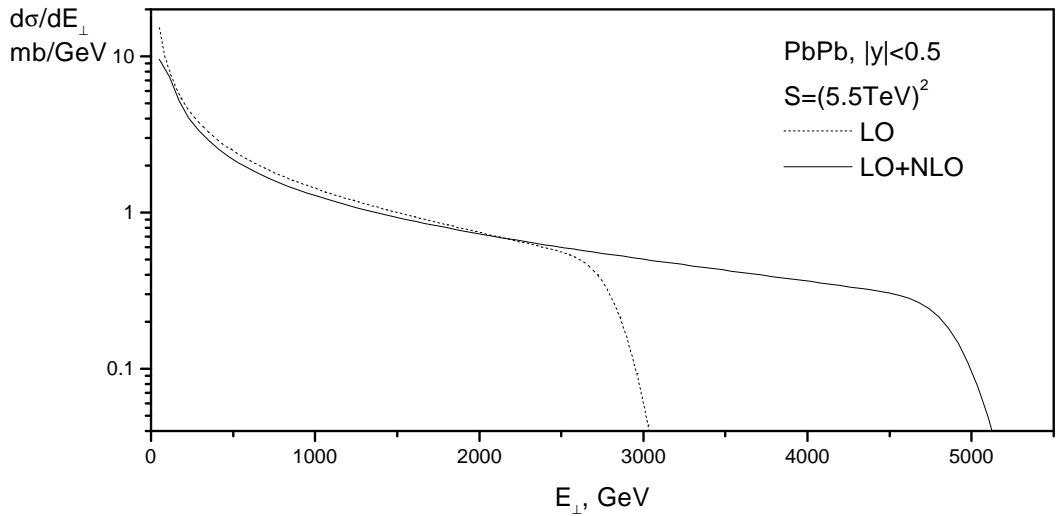,width=14cm}
\end{center}
 \caption{NLO (solid line) and LO (dashed line) transverse energy spectrum in a
unit central rapidity window for PbPb collisions at LHC energy $\sqrt{s}=5500$
GeV}

\label{PbPb5500}
\end{figure}

In this section we turn to an estimate
of the transverse energy production in
 nuclear collisions in the Glauber type approach, in which  they are
considered to be an impact parameter averaged superposition of hadron-hadron
collisions.
We shall follow the geometrical approach similar to that of \cite{EKL} and
consider the
gaussian approximation to the transverse energy distribution at  given impact
parameter in the collision of two nuclei with atomic numbers $A$ and $B$
\begin{equation}
{d \omega_{AB}  \over {d E_\perp}} = {1 \over \sqrt{2 \pi D_{AB}}}
\exp
(-{(E_{\perp} - \langle E_{\perp} \rangle_{AB} (b))^2 \over {2
D_{AB}(b)}}),
\label{ABb}
\end{equation}
where $\langle E_{\perp} \rangle_{AB} (b)$ is a mean transverse energy produced
at given impact parameter $b$ :
\begin{equation}
\langle E_{\perp} \rangle_{AB} (b)=
AB P_{AB}(b) \langle E_{\perp} \rangle_{pp} (b),
\end{equation}
$D_{AB}$ is a dispersion of the $E_{\perp}$ distribution given by
\begin{equation} D_{AB}(b) = AB P_{AB}(b)(\langle E^2_{\perp} \rangle _{pp} -
P_{AB}(b) \langle E_{\perp} \rangle ^2_{pp}), \end{equation}
and $P_{AB}$ is a probability of nuclear scattering at given impact
parameter where the normalization is fixed by
the inelastic cross section for minijet production in
$pp$ collisions Eq.~(\ref{sE0}).
Let us note that the Glauber model we use is similar to
that of \cite{EKL} in that the transverse energy distribution at a given value
of the impact parameter is approximated with the Gaussian, see Eq.~(\ref{ABb}),
but differs from it in using the Bernoulli process instead of Poissoninan one in
\cite{EKL} and the different overall normalization at the integrated minijet
production cross section in the rapidity window under consideration.
The final expression for the cross section of transverse energy production
in nuclear collisions is obtained from Eq.~(\ref{ABb}) by integrating over the
impact parameter
\begin{equation}
 {d \sigma \over d E_\perp} = \int d^2 b\ {d \omega_{AB} \over
{d E_\perp}}
\end{equation}
The resulting transverse energy distributions are plotted in Fig.~\ref{PbPb200}

and Fig.~\ref{PbPb5500}
 for PbPb collisions for RHIC and LHC energies correspondingly. We see that the
main effect of taking into account the NLO corrections is a considerable
broadening of
the shoulder of the distribution which
basically follows form the increase in the magnitude of the transverse
energy production cross section from LO to NLO.
At the same time the height of transverse energy distribution basically does not
change. The above results show that to this order
in perturbative QCD computation the NLO prediction is an increased
event-by-event produced transverse energy providing more favorable conditions
for collective behavior of the produced gluon system, its thermalization, etc.
as compared to the leading order calculations.

\section{Discussion}

The results of the above-presented calculation of the next-to-leading order
corrections to the transverse energy flow into a unit rapidity interval in the
central region show that the NLO corrections to the Born calculation of the
transverse energy distribution in pp collisions
based on the lowest order $2 \to 2$ scattering
are substantial, so that the effective K-factors are $K_{RHIC} \sim 1.9$ and
$K_{LHC} \sim 2.1$, in accordance with the "expected" ones previously used in
the literature \cite {XNW}, \cite{KE1}.

 The cross section of transverse energy production in pp collisions serves as a
basic building block in the geometrical Glauber model of nuclear collisions.
Switching from LO to NLO cross section of $E_{\perp}$ production results
in a substantial broadening of the minijet transverse energy distribution
in nuclear collision
providing more favorable conditions for subsequent collective effects
characteristic for dense parton systems such as quark-gluon plasma to manifest
themselves. Let us note, that the form of this distribution does not change a
lot.

  The effective K-factors being substantial (although not overwhelmingly large)
certainly make it necessary to develop a more accurate treatment of minijet
production. To achieve this goal
 one has to solve two interrelated problems.
Both
are linked to the large value of the integrated minijet
transverse energy production cross section in the leading twist
approximation discussed in the second section and
the resulting conflict
with unitarity at low minijet transverse energies at LHC, at least when
considering large enough rapidity windows.

First of all, one has to get a reliable estimate of the higher order corrections
to the hard blob responsible for $E_{\perp}$ production.
 Most probably it will require resumming the perturbative
contributions to all orders, because even if the hypothetical
next-to-next-to leading order calculation reduces the K-factor, the arising
large cancellations would require further improvement of the accuracy of the
calculation, etc.  Such resummation program has been
successfully performed in the case of jet pair production at the kinematical
threshold \cite{RJ}. Unfortunately at present it is not clear how to extend this
program to the
minijet production kinematics.

The second problem is even more difficult and is related to the
necessity of a reliable computation
of the nonlinear contributions to (mini)jet production which are quite important
at high energies both in hadron and nuclear collisions \cite{XNW} and
photoproduction \cite{PHU}.
The current approach to describing
nonlinear effects is based on the ad hoc
eikonal unitarization scheme, see e.g. \cite{XNW}. This scheme does not take
into account the processes in which several nucleons are involved in transverse
energy production. This problem was recently reanalyzed in \cite{KM} for pA
collisions showing in particular an interesting connection between the
nonlinear effects in the structure functions and those in the spectrum of
emitted gluons. An advanced analysis of the nonlinear effects
at the example of nuclear effects in jet photoproduction
\cite{JWQ} has revealed a number of interesting features possibly relevant
also for the computation of nonlinear effects in hadroproduction of
jets.
One of the most striking aspects of this treatment is that
although diagrammatically the nonlinear effects initially
look as a superposition
of the leading twist contributions,
the final result appears to be a next-to-leading twist
one due to a delicate cancellation between various diagrams
\cite{JWQ}, \cite{KMS}.

  Various theoretical schemes possibly responsible for taming the growth
of the minijet production cross sections were discussed. One of them
is based on accounting for nonlinear effects in quasiclassical approach,
\cite{JKMW}, \cite{GM}. Here at tree level the nonperturbative lattice
calculations of minijet production in nuclear collisions were performed in
\cite{KV}. A more traditional treatment based on accounting for nonlinear
effects in QCD evolution equations is described in the review \cite{LR}.

  Another related problem  is a necessity of switching from the collinear to
high energy factorization in describing the minijet production at high energies,
see e.g.\cite{ELR}, \cite{GM}.

In summary, the computed NLO corrections to the minijet transverse
energy production in hadron and nuclear collisions turned out to be
substantial both for RHIC and LHC energies. Qualitatively this enhances the
energy production in the central region and significantly broadens
the transverse energy spectrum in nuclear collisions. However much
work is still ahead in order to improve the accuracy of these predictions.

\begin{center}
{\it Acknowledgements}
\end{center}
\medskip

We are  grateful to D.~Soper, L.~McLerran, K.~Kajantie, Z.~Kunzst, K.~Eskola,
I.M~Dremin, I.V.~Andreev and S.G.~Matinyan for useful and insightful
discussions.

Special thanks are  to Davison Soper for discussion that initiated the above
work and providing the NLO code and Kari Eskola for reading the manuscript and
useful comments.

A.L. is grateful for hospitality and support at the
Institute for Theoretical Physics, University of Minnesota, where this work had
been started, at CERN Theory Division, where the major part of this work
was done, and at Service de Physique Theorique de Saclay, where it was
finalized.

This work was also partially supported by Russian Fund for Basic Research, Grant
96-02-16210.

\end{document}